\documentclass[a4paper,11pt]{revtex4-1}
\usepackage{graphicx} % standard LaTeX graphics tool;
\usepackage{amsmath} % For getting proper math eqns
\usepackage{amssymb} % Used for getting various symbols eg. gtrsim, lesssim etc.
\usepackage{bm} % For bold math (esp of lower greek letters)
\usepackage{dcolumn}% Align table columns on decimal point
\usepackage{color}
\usepackage{amsfonts}
\usepackage{varioref}
\usepackage{mathrsfs}
\usepackage{graphicx}
\usepackage{subfigure}
\usepackage{latexsym}
\usepackage{amsmath}
\usepackage{multirow}
\usepackage{amssymb}
\usepackage{textcomp}
\usepackage{amsbsy}
\usepackage{graphics}
\usepackage{epstopdf}
\usepackage{color}
\usepackage{natbib}

\RequirePackage[colorlinks,citecolor=blue,urlcolor=magenta,linkcolor=blue]{hyperref}
\input epsf

\allowdisplaybreaks[4]
\begin{document}
\tolerance=5000
\title{Dark energy era with a resolution of Hubble tension in generalized entropic cosmology}

\author{Priyanka~Adhikary$^{1}$\,\thanks{priyankaadhikary35@gmail.com},
Sudipta~Das$^{1}$\,\thanks{sudipta.das@visva-bharati.ac.in},
Sergei~D.~Odintsov$^{2,3}$\,\thanks{odintsov@ieec.uab.es},
Tanmoy~Paul$^{1}$\,\thanks{tanmoy.paul@visva-bharati.ac.in}} \affiliation{
$^{1)}$ Department of Physics, Visva-Bharati University, Santiniketan 731235, India\\
$^{2)}$ ICREA, Passeig Luis Companys, 23, 08010 Barcelona, Spain\\
$^{3)}$ Institute of Space Sciences (ICE, CSIC) C. Can Magrans s/n, 08193 Barcelona, Spain}

%\date{}

\tolerance=5000

\begin{abstract}
We propose a new dark energy (DE) model from four parameter generalized entropy function of apparent horizon in a spatially flat universe. Such kind of generalized entropy is able to generalize all the known entropies proposed so far, for suitable representations of the entropic parameters. It turns out that the scenario can describe the correct thermal history of the universe, with the sequence of matter and dark energy epochs. Comparing with the $\Lambda$CDM model, the proposed generalized entropic DE model provides a higher value of present Hubble parameter for certain range of entropic parameter(s) leading to a possible resolution of Hubble tension issue. We confront the scenario with CC, PantheonPlus+SH0ES, DESI DR1 and compressed Planck likelihood datasets, which clearly depicts the phenomenological viability of the present model for some best fitted values of entropic parameter(s) that are indeed consistent with the resolution of Hubble tension.
\end{abstract}
%%%%%%%%%%%%%%%%%%%%%%%%%%%%%%%%%%%%%%%%%%%%%%%%%%%%%%%%%%%%%%%%%%%%%%%%%%%%%%%%%%%%%%%%%%%%%%%%%%%
%%%%%%%%%%%%%%%%%%%%%%%%%%%%%%%%%%%%%%%%%%%%%%%%%%%%%%%%%%%%%%%%%%%%%%%%%%%%%%%%%%%%%%%%%%%%%%%%%%%
%%%%%%%%%%%%%%%%%%%%%%%%%%%%%%%%%%%%%%%%%%%%%%%%%%%%%%%%%%%%%%%%%%%%%%%%%%%%%%%%%%%%%%%%%%%%%%%%%%%
%\newpage
%%%%%%%%%%%%%%%%%%%%%%%%%%%%%%%%%%%%%%%%%%%%%%%%%%%%%%%%%%%%%%%%%%%%%%%%%%%%%%%%%%%%%%%%%%%%%%%%%%%
%%%%%%%%%%%%%%%%%%%%%%%%%%%%%%%%%%%%%%%%%%%%%%%%%%%%%%%%%%%%%%%%%%%%%%%%%%%%%%%%%%%%%%%%%%%%%%%%%%%
%%%%%%%%%%%%%%%%%%%%%%%%%%%%%%%%%%%%%%%%%%%%%%%%%%%%%%%%%%%%%%%%%%%%%%%%%%%%%%%%%%%%%%%%%%%%%%%%%%%
%\pacs{}

\maketitle

\section{Introduction}\label{SecI}

In thermodynamic description of apparent horizon in cosmology, the entropy of apparent horizon actually fixes the underlying cosmological field equations \cite{Jacobson_1995,Cai:2005ra,Cai:2006rs,Nojiri:2025gkq} (for a recent review on entropic cosmology, see \cite{Nojiri:2024zdu}). It is well known that the Bekenstein-Hawking like entropy of apparent horizon, along with the standard thermodynamic law of the horizon, leads to the usual FLRW equations. Beside the first law of thermodynamics, the natural validation of the second law of thermodynamics in cosmological context has been recently proposed in \cite{Odintsov:2024ipb,Paul:2025rqe}. One of the distinctive features of Bekenstein-Hawking entropy is that it scales as the area of the horizon, unlike to the classical thermodynamics where the entropy goes by the volume of a thermodynamic system under consideration \cite{PhysRevD.7.2333,Hawking:1975vcx}. Owing to such distinctive features of Bekenstein-Hawking entropy, and depending on non-additive statistics, several other forms of entropies have been proposed like --- the Tsallis entropy \cite{Tsallis:1987eu}, the R\'{e}nyi entropy \cite{Renyi}, the Barrow entropy \cite{Barrow_2020}, the Sharma-Mittal entropy \cite{Sayahian_Jahromi_2018}, the Kaniadakis entropy \cite{Kaniadakis_2005}, the Loop Quantum gravity entropy \cite{Majhi:2017zao} etc. Clearly, such non-additive entropies have certain modification(s) than the Bekenstein-Hawking entropy, which in turn have rich cosmological consequences starting from inflation (or bounce) to dark energy era \cite{Bousso_2002, fischler1998, Saridakis:2020zol, Sinha_2020, Adhikary:2021, Nojiri:2021iko, Nojiri:2020wmh}. The interesting point to be noted that all of these entropies share some common properties, like --- (i) they all are monotonically increasing function of Bekenstein-Hawking entropy variable ($S$), and (ii) they tend to zero in the limit of $S \rightarrow 0$. Due to such common features, the immediate question that naturally arises can be recast as follows --- ``Does there exist any generalized form of entropy that can generalize all the known entropies proposed so far ?" In this context some of our authors proposed few parameter dependent generalized entropy function that proves to unify all the known entropies within a single umbrella for suitable representation of generalized entropic parameters \cite{Nojiri:2022aof,Nojiri_2022,Odintsov:2022qnn}. In particular, according to the conjecture stated in \cite{Nojiri_2022}, the minimum number of parameters required for a generalized entropy function is equal to four. However such 4-parameter generalized entropy shows a singular behaviour at the instant when the Hubble parameter vanishes which, for instance, occurs in the context of bouncing cosmology. Consequently a 5-parameter dependent non-singular generalized entropy has been proposed, which proves to be non-singular (even in the context of bouncing cosmology) and at the same time, is able to generalize all the aforementioned entropies \cite{Odintsov:2022qnn}. Thus as a whole, the generalized entropies having four parameters and five parameters (constructed in \cite{Nojiri_2022} and in \cite{Odintsov:2022qnn} respectively) are the minimal constructions of generalized version of entropy depending on whether the universe passes through $H = 0$ during its cosmological evolution (with $H$ being the Hubble parameter of the universe). Some possible implications of generalized entropies to cosmology as well as to black hole physics are discussed in \cite{Nojiri:2022aof,Nojiri_2022,Odintsov:2022qnn,Odintsov_2023,Bolotin:2023wiw,Lymperis:2023prf,doi:10.1142/S0219887824503389,Odintsov:2024sbo,Tyagi:2025zov,Tariq:2025wiy,Odintsov:2025sew}.

Cosmological observations \cite{riess1998observational, perlmutter1999measurements, arnaud2016planck, ahn2012ninth} have also provided us firm evidences in favour of the existence of both dark energy and dark matter components in the universe, but their true nature still remains unknown. The dark energy component, which constitutes about 70\% of the total energy budget of the universe, should exert a large negative pressure in order to counter balance the gravitational force.  A plethora of dark energy models have been constructed, viz, the cosmological constant model ($\Lambda$CDM), dynamical dark energy models \cite{RatraPRD1988, CaldwellPRL2003, BentoPRD2004, Yang_2017, Yang_2019,al2016divergence, mamon2015study}, holographic dark energy model \cite{Bousso_2002, fischler1998, Saridakis:2020zol, Sinha_2020, Adhikary:2021} etc., but no single theory can be definitively considered as the best candidate for dark energy. For instance, the $\Lambda$CDM model perfectly describes the dark energy during the late universe, except the fact that the $\Lambda$CDM model is plagued with the Hubble tension \cite{Verde_2019,Di_Valentino_2021,Jedamzik:2020zmd,Di_Valentino_2021CQG,Kamionkowski:2022pkx,2024Lverde,Knox_2020,Khalife:2023qbu,PERIVOLAROPOULOS2022101659,Abdalla_2022}. The tension arises due to the discrepancy in the measured value of the Hubble parameter $H_0$. The cosmic microwave background data from the Planck satellite along with Baryon Acoustic Oscillation (BAO) data \cite{Planck2020, BAO2017, BAO2011}, Big Bang Nucleosynthesis (BBN) \cite{BBN2021}  and Dark Energy Survey (DES) \cite{DES12018, DES22018, DES:2017tss}  have constrained the value to be $H_0 \sim (67.0 - 68.5)$km/s/Mpc, whereas the measurement from the SH0ES, TRGB and H0LiCOW collaborations \citep{Sh0ES2019, H0LiCOW2019} predict the value to be $H_0 \sim (74.03 \pm 1.42)$ km/s/Mpc. This statistical anomaly in the measurement for these two types of data could not be addressed by $\Lambda$CDM model \citep{Abdalla:2022yfr} and is an open problem in cosmology. Thus the true nature of dark energy still remains one of the biggest mysteries in cosmology and the search is still on. A large number models have been proposed to address the ``Hubble Tension" \cite{2016PhLB_Valentino, Di_Valentino_2020, Vagnozzi_2020, Alestas_2020, Di_Valentino_2019,2021PRD_Banerjee}. Both early universe \cite{Poulin_2019, Kamionkowski:2022pkx, Poulin:2023lkg, Niedermann_2022, Vagnozzi_2021} as well as late universe solutions \cite{Yang_2020,Alestas2021b,Vagnozzi:2023nrq,Odintsov:2020qzd,Pedrotti:2024kpn,Elizalde:2024rvg,Yang:2018uae,Mamon_2017,Niedermann:2020dwg,DiValentino:2019jae,DiValentino:2020naf,ROY2022101037,Verde_2019,Di_Valentino_2021,Jedamzik:2020zmd,Di_Valentino_2021CQG,Kamionkowski:2022pkx,2024Lverde,Knox_2020,Khalife:2023qbu,Odintsov:2025sew} have been proposed in different studies. For a detailed review, one can look at \cite{cosmoversewhitepaper, HubbleTensionbook2024}.

In the present paper, we will focus on the dark energy era from the perspective of thermodynamic cosmology, based on the thermodynamic law of apparent horizon. In particular, we will consider the 4-parameter generalized entropy in the present paper to investigate its possible consequences towards dark energy era of the universe. The 4-parameter generalized entropy is motivated due to the following reasons: (a) it is the minimal version of generalized entropy functions, and (b) it leads to certain deviation(s) from the usual $\Lambda$CDM model, that may give rise to a natural resolution of Hubble tension. 

Our paper is organized as follows: in Sec.~\ref{SecII}, we will discuss the basic formalism of apparent horizon thermodynamics and will determine the cosmological field equations corresponding to a
general form of horizon entropy. The corresponding field equations for four parameter generalized entropy and the associated entropic energy density is discussed in Sec.~\ref{Sec-III}. The next section, Sec.~\ref{sec-DE}, deals with the relevant cosmological parameters for the dark energy model and the possible resolution of Hubble tension from the perspective of four parameter generalized entropy function. The data analysis and results are demonstrated in Sec.~\ref{results}, and some concluding remarks will be presented in Sec.~\ref{Sec-conclusion}.

\section{Thermodynamics of apparent horizon and cosmological field equations}\label{SecII}

We consider the $(3+1)$ dimensional spatially flat Friedmann-Lema\^{i}tre-Robertson-Walker (FLRW) universe, whose metric is given by,
\begin{align}
\label{dS7}
ds^2 = \sum_{\mu,\nu=0,1,2,3} g_{\mu\nu} dx^\mu dx^\nu = - dt ^2 + a( t )^2 \left( d r ^2 + r ^2 {d\Omega_{2}}^2 \right) \, ,
\end{align}
where ${d\Omega_{2}}^2$ is the line element of a $2$ dimensional sphere of unit radius (particularly on the surface of the sphere).
We also define
\begin{align}
\label{dS7B}
d{s_\perp}^2 = \sum_{M,N=0,1} h_{\mu\nu} dx^M dx^N = - dt ^2 + a( t )^2 d r ^2 \, .
\end{align}
The radius of the apparent horizon $R_\mathrm{h}=R\equiv a(t)r$ for the FLRW universe is given by the solution of the equation 
$h^{MN} \partial_M R \partial_N R = 0$ (see \cite{Cai:2005ra}) which immediately leads to,
\begin{align}
\label{dS14A}
R_\mathrm{h}=\frac{1}{H}\, ,
\end{align}
with $H\equiv \frac{1}{a}\frac{da}{d t }$ representing the Hubble parameter of the universe. In the context of entropic cosmology, the apparent horizon is considered to have a thermal behaviour, and one may motivate this by following arguments:
\begin{itemize}
 \item The matter fields inside the horizon exhibits a flux from inside to outside of the horizon, and the flux is outward (or inward) during the accelerating (or decelerating) stage of the universe. In the case of outward flux, the amount of matter fields inside the horizon decrease, which leads to a decrease of matter fields' entropy. This violates the second law of thermodynamics which states that the change of entropy should be positive. Therefore in order to validate the second law of thermodynamics in cosmological context, we need to consider an entropy of the apparent horizon, which can manage the change of total entropy to be positive (here it is important to mention that the total entropy is defined by ``horizon entropy'' + ``matter fields' entropy'').

 \item The FLRW equations are symmetric under $t \rightarrow -t$ (time reversal symmetric), and thus it always come with a contracting solution along with the expanding one. However the observational data indicates that the universe is expanding. Hence the natural question that arises is --- ``why does the universe choose the expanding solution?'' To answer this question, we may associate a thermal behaviour to the apparent horizon. Then the second law of thermodynamics actually disagrees the contracting solution in order to maintain a positive change of total entropy.
\end{itemize}
Owing to thermal behaviour, the apparent horizon follows the thermodynamic law \cite{Cai:2005ra,Cai:2006rs,Nojiri:2023wzz}:
\begin{eqnarray}
 T_\mathrm{h}dS_\mathrm{h} = -dE + WdV\, ,
\label{law-1}
\end{eqnarray}
where $S_\mathrm{h}$ is the entropy of the apparent horizon and $T_\mathrm{h}$ is the corresponding temperature which is fixed by the surface gravity of the horizon. In the case of a spatially flat, homogeneous and isotropic universe, $T_\mathrm{h}$ is given by,
\begin{eqnarray}
 \label{AH2}
T_\mathrm{h} = \frac{H}{2\pi} \left| 1 + \frac{\dot{H}}{2H^2} \right|\, .
\end{eqnarray}
Moreover $E = \rho V$ is the total energy of matter fields inside the horizon and $W = \frac{1}{2}\left(\rho - p\right)$ is the work density by the matter fields ($\rho$ and $p$ symbolize the energy density and the pressure of the matter fields). Equation~(\ref{law-1}) clearly argues that the change of horizon entropy causes from two reasons --- (a) by the term $-dE$ which represents the amount of matter fields that decrease inside the horizon within a time $dt$, and (b) by the term $WdV$ representing the work done by the matter fields at time $dt$. Therefore both the terms actually indicate an effective outward flux of the matter fields from inside to outside of the horizon. Since the horizon divides the observable universe from the unobservable one, such outflux of the matter fields can be thought as some information loss of the observable universe, which in turn gives rise to an entropy of the apparent horizon. 

The thermodynamics of the apparent horizon governed by equation~(\ref{law-1}) fixes the cosmological field equations, and depending on the form of $S_\mathrm{h}$, the field equations get modified. However irrespective of the form, $S_\mathrm{h}$ shares some common properties like :
\begin{itemize}
 \item $S_\mathrm{h}$ is a monotonic increasing function of the Bekenstein-Hawking entropy variable $S = A/(4G)$ (where $A = 4\pi R_\mathrm{h}^2$ denotes the area of the apparent horizon),

 \item $S_\mathrm{h}$ goes to zero in the limit of $S \rightarrow 0$, which can be thought as equivalent of the third law of thermodynamics.
\end{itemize}
In the following, we derive the gravitational field equations from equation~(\ref{law-1}) for a general form of the horizon entropy given by $S_\mathrm{h}$. Taking $E = \rho V$ and $W = \frac{1}{2}\left(\rho - p\right)$ into account, equation~(\ref{law-1}) can be written by,
\begin{eqnarray}
 T_\mathrm{h}\dot{S}_\mathrm{h} = -\dot{\rho}V - \frac{1}{2}\left(\rho + p\right)\dot{V}~~,
 \label{HE-1}
\end{eqnarray}
where the overdot symbolizes $\frac{d}{dt}$ of the respective quantity. Considering the conservation equation of the matter fields, i.e.
\begin{eqnarray}
 \dot{\rho} + 3H\left(\rho + p\right) = 0~~,
 \label{conservation-matter}
\end{eqnarray}
Equation~(\ref{HE-1}) can be expressed as,
\begin{eqnarray}
 T_\mathrm{h}\dot{S}_\mathrm{h} = \left(\rho + p\right)\left\{3HV - \frac{\dot{V}}{2}\right\}~~,
 \label{HE-2}
\end{eqnarray}
which, owing to $V=\frac{4}{3}\pi R_\mathrm{h}^3$, takes the following form:
\begin{eqnarray}
 \dot{S}_\mathrm{h} = \frac{8\pi}{H^3}\left(\rho + p\right)~~.
 \label{HE-3}
\end{eqnarray}
As mentioned above that $S_\mathrm{h}$ is a function of the Bekenstein-Hawking entropy variable $S$, and thus equation~(\ref{HE-3}) can be expressed by,
\begin{eqnarray}
 \dot{H}\left(\frac{\partial S_\mathrm{h}}{\partial S}\right) = -4\pi G\left(\rho + p\right)~~,
 \label{HE-4}
\end{eqnarray}
where we have used $S = \frac{\pi}{GH^2}$ is the Bekenstein-Hawking entropy and $\dot{S} = -\frac{2\pi}{G}\left(\frac{\dot{H}}{H^3}\right)$. Integration of both sides of equation~(\ref{HE-4}) along with the energy conservation of the matter fields yields the following expression:
\begin{eqnarray}
 \int \left(\frac{\partial S_\mathrm{h}}{\partial S}\right) d\left(H^2\right) = \frac{8\pi G}{3}\rho + \frac{\Lambda}{3}~~,
 \label{HE-5}
\end{eqnarray}
where $\Lambda$ is the constant of integration (also known as the cosmological constant), and the integration can be performed once we consider a specific form of the horizon entropy in terms of the Bekenstein-Hawking entropy variable (i.e. $S_\mathrm{h} = S_\mathrm{h}(S)$).

Equation~(\ref{HE-5}) and equation~(\ref{HE-4}) act as the first and second Friedmann equation respectively in the horizon cosmology for a general form of the horizon entropy. It may be noted that for $S_\mathrm{h} = S$, both of these equations reduce to the usual Friedmann equations of Einstein gravity. Thus the entropic cosmology with the Bekenstein-Hawking horizon entropy is similar to that in case of Einstein gravity, otherwise, some other form of the horizon entropy will result to a modified Friedmann equations. In the present context, we will consider $S_\mathrm{h}$ to be of the form of 4-parameter generalized entropy and will focus on the dark energy era of the universe.

\section{Modified Friedmann equations corresponding to the 4-parameter generalized entropy}\label{Sec-III}
The 4-parameter generalized entropy is given by \cite{Nojiri_2022},
\begin{eqnarray}
 S_\mathrm{g}\left[\alpha_+,\alpha_-,\beta,\gamma \right] = \frac{1}{\gamma}\left[\left(1 + \frac{\alpha_+}{\beta}~S\right)^{\beta}
 - \left(1 + \frac{\alpha_-}{\beta}~S\right)^{-\beta}\right] \, ,
\label{gen-entropy}
\end{eqnarray}
which is able to generalize the known entropies like the Tsallis entropy, the R\'{e}nyi entropy, the Barrow entropy, the Sharma-Mittal entropy, the Kaniadakis entropy and the Loop Quantum Gravity entropy at suitable limit of the parameters. With the above form of horizon entropy, i.e. $S_\mathrm{h} \equiv S_\mathrm{g}$, equation~(\ref{HE-4}) and equation~(\ref{HE-5}) take the following form,
\begin{align}
\frac{1}{\gamma}\left[\alpha_{+}\left(1 + \frac{\pi \alpha_+}{\beta GH^2}\right)^{\beta - 1}
+ \alpha_-\left(1 + \frac{\pi \alpha_-}{\beta GH^2}\right)^{-\beta-1}\right]\dot{H} = -4\pi G\left(\rho + p\right)
\label{FRW-1}
\end{align}
and
\begin{align}
\frac{GH^4\beta}{\pi\gamma}&\,\left[ \frac{1}{\left(2+\beta\right)}\left(\frac{GH^2\beta}{\pi\alpha_-}\right)^{\beta}~
2F_{1}\left(1+\beta, 2+\beta, 3+\beta, -\frac{GH^2\beta}{\pi\alpha_-}\right) \right. \nonumber\\
&\, \left. + \frac{1}{\left(2-\beta\right)}\left(\frac{GH^2\beta}{\pi\alpha_+}
\right)^{-\beta}~2F_{1}\left(1-\beta, 2-\beta, 3-\beta, -\frac{GH^2\beta}{\pi\alpha_+}\right) \right] = \frac{8\pi G\rho}{3} + \frac{\Lambda}{3} \,,
\label{FRW-2}
\end{align}
respectively, where we use $S = A/(4G) = \pi/(GH^2)$ and $2F_1(\mathrm{arguments})$ denotes the Hypergeometric function. Equation~(\ref{FRW-1}) and equation~(\ref{FRW-2}) represent the modified Friedmann equations corresponding to the generalized entropy function $S_\mathrm{g}$. The above two equations can be equivalently written by,
\begin{align}
\dot{H}=&\,-4\pi G\left[\left(\rho + \rho_\mathrm{g}\right) + \left(p + p_\mathrm{g}\right)\right] \,,\nonumber\\
H^2=&\, \frac{8\pi G}{3}\left(\rho + \rho_\mathrm{g}\right) + \frac{\Lambda}{3} \, ,
\label{final FRW}
\end{align}
where $\rho_\mathrm{g}$ and $p_\mathrm{g}$ are given by,
\begin{align}
\rho_\mathrm{g} = \frac{3}{8\pi G}\left\{ H^2 - \frac{GH^4\beta}{\pi\gamma}\right. &\,\left[ \frac{1}{\left(2+\beta\right)}
\left(\frac{GH^2\beta}{\pi\alpha_-}\right)^{\beta}~2F_{1}\left(1+\beta, 2+\beta, 3+\beta, -\frac{GH^2\beta}{\pi\alpha_-}\right) \right. \nonumber\\
& \left. \left. + \frac{1}{\left(2-\beta\right)}\left(\frac{GH^2\beta}{\pi\alpha_+}\right)^{-\beta}~2F_{1}\left(1-\beta, 2-\beta, 3-\beta, -\frac{GH^2\beta}{\pi\alpha_+}\right)
\right] \right\} \,,
\label{efective energy density}
\end{align}
and
\begin{align}
p_\mathrm{g} = \frac{\dot{H}}{4\pi G}\left\{\frac{1}{\gamma}\left[\alpha_{+}\left(1 + \frac{\pi \alpha_+}{GH^2\beta}\right)^{\beta - 1}
+ \alpha_-\left(1 + \frac{\pi \alpha_-}{GH^2\beta}\right)^{-\beta-1}\right] - 1\right\} - \rho_\mathrm{g} \,,
\label{effective pressure}
\end{align}
respectively. The $\rho_\mathrm{g}$ and $p_\mathrm{g}$ are termed as entropic energy density and entropic pressure corresponding to the 4-parameter generalized entropy. In order to have a better understanding, let us consider the condition: $GH^2 \ll 1$ which is indeed a valid consideration as we will eventually deal with the late time era of the universe where the Hubble parameter safely satisfies this condition. Such a condition leads to a simplified form of $\rho_\mathrm{g}$ and $p_\mathrm{g}$ as,
\begin{eqnarray}
 \rho_\mathrm{g}&=&\frac{3H^2}{8\pi G}\left[1 - \frac{\alpha_{+}}{\gamma(2-\beta)}\left(\frac{GH^2\beta}{\pi\alpha_{+}}\right)^{1-\beta}\right] \, ,\nonumber\\
 p_\mathrm{g}&=&-\frac{\dot{H}}{4\pi G}\left[1 - \frac{\alpha_{+}}{\gamma}\left(\frac{GH^2\beta}{\pi\alpha_{+}}\right)^{1-\beta}\right] - \rho_\mathrm{g} \, ,
 \label{simplified rho-p-g}
\end{eqnarray}
where we retain the leading order term in the Taylor series of the Hypergeometric function. Equation~(\ref{simplified rho-p-g}) clearly indicates that $\rho_\mathrm{g}$ and $p_\mathrm{g}$ identically vanish for the choice of the entropic parameters: $\alpha_{+} = \gamma$ and $\beta = 1$ --- such a choice reduces the 4-parameter generalized entropy ($S_\mathrm{g}$) to the Bekenstein-Hawking entropy ($S$). Thus one may argue that $\rho_\mathrm{g}$ and $p_\mathrm{g}$ actually encapsulate the information of the generalized entropy over the Bekenstein-Hawking one, or equivalently, the $\rho_\mathrm{g}$ and $p_\mathrm{g}$ contain the modification of the FLRW equations over the case of usual Einstein gravity. In the next section, we aim to study the cosmological implications of the modified Friedmann equation~(\ref{final FRW}) during the dark energy era.

\section{Dark energy era and a possible resolution of Hubble tension}\label{sec-DE}

In this section we will concentrate on late time cosmological implications of the generalized entropy function ($S_\mathrm{g}$). The main equations will be equation~(\ref{final FRW}) where the entropic energy density ($\rho_\mathrm{g}$) and the cosmological constant contribute to the dark energy density, while $\rho$ acts as the energy density for the pressureless dust and the radiation fluid, i.e. $\rho = \rho_\mathrm{m} + \rho_\mathrm{R}$. However due to $\rho_\mathrm{R} \ll \rho_\mathrm{m}$ at late time, we take $\rho \approx \rho_\mathrm{m}$. In particular, the dark energy density and the dark pressure are given by,
\begin{eqnarray}
 \rho_\mathrm{D}=\rho_\mathrm{g} + \frac{3}{8\pi G}\left(\frac{\Lambda}{3}\right) = \frac{3}{8\pi G}\left[\frac{\Lambda}{3} + H^2\left(1 - \frac{\sigma_\mathrm{0}}{(2-\beta)}H^{2(1-\beta)}\right)\right]
 \label{de-0}
 \end{eqnarray}
 and
\begin{eqnarray}
\rho_\mathrm{D} + p_\mathrm{D} = \rho_\mathrm{g} + p_\mathrm{g} = -\frac{\dot{H}}{4\pi G}\left[1 - \sigma_\mathrm{0}H^{2(1-\beta)}\right]~~,
 \label{de-1}
\end{eqnarray}
where we use $\rho_\mathrm{g}$ and $p_\mathrm{g}$ from equation (\ref{simplified rho-p-g}), and recall that the cosmological constant has the equation of state like $\rho_\mathrm{\Lambda} + p_\mathrm{\Lambda} = 0$. Moreover $\sigma_0$ (having mass dimension of $[M^{-2(1-\beta)}]$) is symbolized by (in terms of the entropic parameters)
\begin{eqnarray}
 \sigma_\mathrm{0} = \frac{\alpha_{+}}{\gamma}\left(\frac{G\beta}{\pi \alpha_{+}}\right)^{1-\beta}~~.
 \label{sigma}
\end{eqnarray}
The presence of $\rho_\mathrm{g}$ and $p_\mathrm{g}$ in equation~(\ref{final FRW}) actually makes the present scenario different than the standard $\Lambda$CDM scenario. In the context of entropic cosmology, the $\Lambda$CDM model is equivalent to the case when the apparent horizon has the Bekenstein-Hawking like entropy. We will show that a different form of horizon entropy than the Bekenstein-Hawking one, in particular with the 4-parameter generalized entropy of the apparent horizon, not only provides a viable dark energy era of the universe but also resolves the Hubble tension issue, unlike to the $\Lambda$CDM model which is generally plagued with the Hubble tension. Here we would like to further mention that without the cosmological constant ($\Lambda$), the $\rho_\mathrm{g}$ (and $p_\mathrm{g}$) alone proves unable to describe a viable late time era of the universe; in particular --- without the term $\Lambda$ in equation~(\ref{final FRW}), the present scenario with 4-parameter generalized entropy leads to a constant deceleration parameter and thus fails to describe the $transition$ from deceleration (matter dominated era) to the late acceleration of the universe. Therefore both $\Lambda$ as well as $\rho_\mathrm{g}$ (and $p_\mathrm{g}$) are required to have a viable dark energy era of the universe. We will come back to this point in details at some stage.

Equation~(\ref{de-0}) and equation~(\ref{de-1}) immediately leads to the dark energy equation of state (EoS) parameter as follows,
\begin{align}
\omega_\mathrm{D} = p_\mathrm{D}/\rho_\mathrm{D} = -1 - \frac{2\dot{H}\left(1 - \sigma_\mathrm{0}H^{2(1-\beta)}\right)}{\Lambda + 3H^2\left(1 - \frac{\sigma_\mathrm{0}}{(2-\beta)}H^{2(1-\beta)}\right)} \,.
\label{eos-1}
\end{align}
Clearly for $\alpha_{+}=\gamma$ and $\beta = 1$, one gets $\omega_\mathrm{D}=-1$. This is however expected because for such a choice, i.e. for $\alpha_{+}=\gamma$ and $\beta = 1$, the 4-parameter generalized entropy reduces to the Bekenstein-Hawking entropy which results to the $\Lambda$CDM model where the dark energy EoS parameter is $=-1$. However a different choice of the entropic parameters leads to a variable $\omega_\mathrm{D}$ (with respect to the cosmic evolution) that will eventually become important in the present context. With these ingredients in hand, the Friedmann equations are written as,
\begin{align}
H^2 = \frac{8\pi G}{3}\left(\rho_m + \rho_\mathrm{D}\right) \,,\nonumber\\
\dot{H} = -4\pi G\left[\rho_m + \left(\rho_\mathrm{D} + p_\mathrm{D}\right)\right] \,,
\label{FRW-late time-1}
\end{align}
where $\rho_\mathrm{D}$, $p_\mathrm{D}$ has been defined earlier and $\rho_\mathrm{m}$ represents the energy density for the dust fluid. Here we consider that the dust and the dark energy are non-interacting, and thus the individual sector obeys their conservation relation, namely
\begin{align}
\dot{\rho}_\mathrm{m} + 3H\rho_\mathrm{m} = 0 \,,\nonumber\\
\dot{\rho}_\mathrm{D} + 3H\rho_\mathrm{D}\left(1 + \omega_\mathrm{D}\right) = 0 \,.
\label{conservation relation}
\end{align}
Finally we introduce the fractional energy density as,
\begin{align}
\Omega_\mathrm{m} = \left(\frac{8\pi G}{3H^2}\right)\rho_\mathrm{m} \,, \quad
\Omega_\mathrm{D} = \left(\frac{8\pi G}{3H^2}\right)\rho_\mathrm{D} \,,
\label{fractional energy density}
\end{align}
respectively, with $\Omega_\mathrm{m} + \Omega_\mathrm{D} = 1$. Equation~(\ref{conservation relation}) gives $\rho_\mathrm{m} = \rho_\mathrm{m0}\left(\frac{a_\mathrm{0}}{a}\right)^3$ where $a_\mathrm{0}$ is the present time scale factor and $\rho_\mathrm{m0}$ being the matter energy density at $a_\mathrm{0}$ (in the following, the subscript``0'' with a quantity denotes the value of the respective quantity at present time). Owing to this evolution of $\rho_\mathrm{m}$, equation~(\ref{fractional energy density}) leads to $\Omega_\mathrm{m} = \Omega_\mathrm{m0}a_\mathrm{0}^3H_0^2/(a^3H^2)$ which, along with the relation $\Omega_\mathrm{m} + \Omega_\mathrm{D} = 1$, results to the Hubble parameter in terms of the redshift factor ($z$) as follows,
\begin{align}
H(z) = \frac{H_\mathrm{0}\sqrt{\Omega_\mathrm{m0}(1+z)^3}}{\sqrt{1-\Omega_\mathrm{D}(z)}} \,,
\label{Hubble late time}
\end{align}
where $z = \frac{a_\mathrm{0}}{a} - 1$ is the red shift factor (the present time is designated by $z = 0$) and $\Omega_\mathrm{m0} = \rho_\mathrm{m0}\left(8\pi G\right)/\left(3H_\mathrm{0}^2\right)$. Differentiating both sides of equation~(\ref{Hubble late time}) with respect to cosmic time, we obtain the following expression,
\begin{eqnarray}
 \dot{H} = -\frac{H^2}{2\left(1 - \Omega_\mathrm{D}(z)\right)}\left[3\left(1 - \Omega_\mathrm{D}(z)\right) + (1+z)\frac{d\Omega_\mathrm{D}}{dz}\right]~~.
 \label{derivative Hubble late time}
\end{eqnarray}
To arrive at the above expression, the identity $\dot{z} = -H(1+z)$ has been used. By using equation~(\ref{derivative Hubble late time}), we get the dark energy EoS parameter from equation~(\ref{eos-1}) as follows (in terms of the redshift factor),
\begin{eqnarray}
 \omega_\mathrm{D}(z) = -1 + \frac{\mathscr{N}}{\mathscr{D}}\, ,
 \label{eos-final}
\end{eqnarray}
with $\mathscr{N}$ and $\mathscr{D}$ have the following forms,
\begin{eqnarray}
 \mathscr{N} = \left[1 + \frac{(1+z)}{3\big(1 - \Omega_\mathrm{D}(z)\big)}\frac{d\Omega_\mathrm{D}}{dz}\right]
 \left[1 - \sigma_\mathrm{0}\left\{\frac{H_\mathrm{0}^2\Omega_\mathrm{m0}(1+z)^3}{1-\Omega_\mathrm{D}(z)}\right\}^{1-\beta}\right]\, ,
 \label{N}
\end{eqnarray}
and
\begin{eqnarray}
 \mathscr{D} = 1 + \frac{\Lambda\big(1 - \Omega_\mathrm{D}(z)\big)}{3H_\mathrm{0}^2\Omega_\mathrm{m0}(1+z)^3} - \frac{\sigma_\mathrm{0}}{2-\beta}\left\{\frac{H_\mathrm{0}^2\Omega_\mathrm{m0}(1+z)^3}{1-\Omega_\mathrm{D}(z)}\right\}^{1-\beta}\, ,
 \label{D}
\end{eqnarray}
respectively. Furthermore it will be useful to determine the deceleration parameter defined by: $q(z) = -1-\dot{H}/H^2$ which, due to equation~(\ref{derivative Hubble late time}), comes as,
\begin{eqnarray}
 q(z) = -1 + \frac{1}{2\big(1 - \Omega_\mathrm{D}(z)\big)}\left\{3\big(1 - \Omega_\mathrm{D}(z)\big) + (1+z)\frac{d\Omega_\mathrm{D}}{dz}\right\}\, .
 \label{deceleration parameter}
\end{eqnarray}
It is evident that both the dark energy EoS parameter and the deceleration parameter depend on $\Omega_\mathrm{D}(z)$ and its derivative. Therefore in order to have $\omega_\mathrm{D} = \omega_\mathrm{D}(z)$ and $q = q(z)$, i.e. in terms of redshift factor, we need to determine the dark energy density parameter in terms of $z$. For this purpose, by plugging the expression of $\rho_\mathrm{D}$ from equation~(\ref{de-0}) into equation~(\ref{fractional energy density}) and with a little bit of simplification, we get,
\begin{eqnarray}
 \Omega_\mathrm{D}(z) = 1 - H_\mathrm{0}^2\Omega_\mathrm{m0}(1+z)^3\left\{\left(\frac{2-\beta}{\sigma_\mathrm{0}}\right)\left[\frac{\Lambda}{3} + H_\mathrm{0}^2\Omega_\mathrm{m0}(1+z)^3\right]\right\}^{1/(\beta-2)}~~.
 \label{fractional DE late time-1}
\end{eqnarray}
The above expression at $z=0$ provides a constraint relation between $(\Lambda,\Omega_{m0},H_0)$, and is given by,
\begin{eqnarray}
 \Lambda = \left(\frac{3\sigma_\mathrm{0}}{2-\beta}\right)H_\mathrm{0}^{2(2-\beta)} - 3H_\mathrm{0}^2\Omega_\mathrm{m0} \, .
 \label{constraint}
\end{eqnarray}
A differentiation of equation~(\ref{fractional DE late time-1}) with respect to $z$ leads to (after a bit of simplification),
\begin{eqnarray}
 \frac{d\Omega_\mathrm{D}}{dz} = \frac{\left\{3(\beta-1)H_\mathrm{0}^2\Omega_\mathrm{m0}(1+z)^3 + \Lambda(\beta-2)\right\}
 \left\{\frac{2-\beta}{\sigma_\mathrm{0}}\left(1 + \frac{\Lambda}{3H_\mathrm{0}^2\Omega_\mathrm{m0}(1+z)^3}\right)\right\}^{(3-\beta)/(\beta-2)}}
 {\sigma_\mathrm{0}(1+z)\left\{H_\mathrm{0}^2\Omega_\mathrm{m0}(1+z)^3\right\}^{1/(2-\beta)}}\, .
 \label{derivative fractional DE}
\end{eqnarray}
Thus as a whole, equations~(\ref{eos-final}), (\ref{deceleration parameter}) and (\ref{fractional DE late time-1}) provide the analytic expressions of dark energy EoS parameter, deceleration parameter and the dark energy density parameter respectively (in terms of redshift factor), in a spatially flat universe filled with dust matter. It may be realized that $\omega_\mathrm{D}(z)$, $q(z)$ and $\Omega_\mathrm{D}(z)$ depend on the entropic parameters $\beta$ and $\sigma_\mathrm{0}$ along with observationally determined parameters $H_\mathrm{0}$ and $\Omega_\mathrm{m0}$ (recall the constraint relation (\ref{constraint}) that relates $\Lambda$ with $H_\mathrm{0}$ and $\Omega_\mathrm{m0}$).

Before going to the observational details, let us mention the importance of the individual terms $\Lambda$ (cosmological constant) and $\rho_\mathrm{g}$ (entropic energy density) contributed to the dark energy density (see equation~(\ref{de-0})) in the present scenario.
\begin{itemize}
 \item {\underline{Importance of the cosmological constant}}: Without the term $\Lambda$ in equation~(\ref{final FRW}), the dark energy density parameter from equation~(\ref{fractional DE late time-1}) becomes,
 \begin{eqnarray}
  \Omega_\mathrm{D}(z) = 1 - H_\mathrm{0}^2\Omega_\mathrm{m0}(1+z)^3\left\{\left(\frac{2-\beta}{\sigma_\mathrm{0}}\right)\left[H_\mathrm{0}^2\Omega_\mathrm{m0}(1+z)^3\right\}\right]^{1/(\beta-2)}~~.
 \label{fractional DE late time-2}
 \end{eqnarray}
which, at $z=0$ gives the following constraint relation between $\left\{\Omega_\mathrm{m0},H_\mathrm{0}\right\}$:
\begin{eqnarray}
 \Omega_\mathrm{m0} = \frac{\sigma_\mathrm{0}}{(2-\beta)}H_\mathrm{0}^{2-2\beta}\, .
 \label{constraint-2}
\end{eqnarray}
Plugging back the above expression into equation~(\ref{fractional DE late time-2}), one finally gets a simplified form of $\Omega_\mathrm{D}(z)$ in absence of $\Lambda$ as follows,
\begin{eqnarray}
 \Omega_\mathrm{D}(z) = 1 - \Omega_\mathrm{m0}(1+z)^{3(\beta-1)/(\beta-2)}\, .
 \label{fractional DE late time-3}
\end{eqnarray}
Based on this $\Omega_\mathrm{D}(z)$, the deceleration parameter turns out to be a $constant$ (with respect to $z$), in particular,
\begin{eqnarray}
 q = -1 + \frac{1}{2\big(1 - \Omega_\mathrm{D}(z)\big)}\left\{3\big(1 - \Omega_\mathrm{D}(z)\big) + (1+z)\frac{d\Omega_\mathrm{D}}{dz}\right\} = \frac{1-2\beta}{2\beta-4}\, ,
\end{eqnarray}
which clearly fails to describe the $transition$ of the universe from the matter dominated era (deceleration) to the late time acceleration era. Therefore as mentioned earlier that without the term $\Lambda$, the present scenario with $\rho_\mathrm{g}$ (and $p_\mathrm{g}$) alone fails to describe a viable late time cosmic evolution of the universe.

\item {\underline{Importance of the entropic energy density}}: Here we will try to motivate that the present scenario corresponding to the 4-parameter generalized entropy may resolve the Hubble tension, unlike to the standard $\Lambda$CDM cosmology. We start from the fact that the Hubble function in $\Lambda$CDM is given by the relation,
\begin{eqnarray}
 \widetilde{H}(z) = \widetilde{H}_\mathrm{0}\sqrt{\widetilde{\Omega}_\mathrm{m0}(1+z)^3 + 1 - \widetilde{\Omega}_\mathrm{m0}}\, ,
 \label{HT-1}
\end{eqnarray}
where the quantities in $\Lambda$CDM are denoted by a tilde. On other side, the Hubble parameter in the present context of generalized entropic cosmology is given by equation~(\ref{Hubble late time}) with $\Omega_\mathrm{D}(z)$ shown in equation~(\ref{fractional DE late time-1}). In order to demonstrate how the generalized entropic scenario deviates from $\Lambda$CDM at present time, we consider that the $H(z)$ (based on the present model) coincides with $\widetilde{H}(z)$ around the recombination epoch $z = z_\mathrm{rec} \approx 1100$, namely $H(z\rightarrow z_\mathrm{rec}) \approx \widetilde{H}(z\rightarrow z_\mathrm{rec})$, but give $H(z\rightarrow 0) > \widetilde{H}(z\rightarrow 0)$ in order to get a higher value of present Hubble parameter compared to the $\Lambda$CDM case. Now, for the $\Lambda$CDM case,
\begin{eqnarray}
 \widetilde{H}(z \rightarrow z_\mathrm{rec} \gg 1) \approx \widetilde{H}_\mathrm{0}\sqrt{\widetilde{\Omega}_\mathrm{m0}}~z_\mathrm{rec}^{3/2}\, ,
 \label{HT-2}
\end{eqnarray}
while, the Hubble parameter around $z_\mathrm{rec}$ in the present entropic scenario comes as,
\begin{eqnarray}
 H(z \rightarrow z_\mathrm{rec} \gg 1) \approx \left(\frac{2-\beta}{\sigma_\mathrm{0}}\right)^{1/(4-2\beta)}\big[H_\mathrm{0}\sqrt{\Omega_\mathrm{m0}} ~z_\mathrm{rec}^{3/2}\big]^{1/(2-\beta)}\, ,
 \label{HT-3}
\end{eqnarray}
where we use $\Omega_\mathrm{D}(z)$ from equation~(\ref{fractional DE late time-1}). The comparison of the above two equations, i.e. $H(z\rightarrow z_\mathrm{rec}) = \widetilde{H}(z\rightarrow z_\mathrm{rec})$, leads to
\begin{eqnarray}
 H_\mathrm{0} = \frac{1}{\sqrt{\Omega_\mathrm{m0}}}\left(\frac{\sigma_\mathrm{0}}{2-\beta}\right)^2\bigg[\widetilde{H}_\mathrm{0}\sqrt{\widetilde{\Omega}_\mathrm{m0}}\bigg]^{2-\beta}z_\mathrm{rec}^{3(1-\beta)/2}\, ,
 \label{HT-4}
\end{eqnarray}
which in turn provides the present time Hubble parameter for our model in terms of $\Omega_\mathrm{m0}$, entropic parameters ($\beta$, $\sigma_\mathrm{0}$) and the $\Lambda$CDM quantities. The important point to be noted from equation~(\ref{HT-4}) is that for $\beta=1$ and $\sigma_\mathrm{0} =1$, $H_\mathrm{0}$ and $\Omega_\mathrm{m0}$ match with the $\Lambda$CDM cosmology --- this is expected as the 4-parameter generalized entropy reduces to the Bekenstein-Hawking one for such choice of entropic parameters, which in turn leads to the standard $\Lambda$CDM scenario. However for a different choice of entropic parameters, i.e. $\beta \neq 1$ and (or) $\sigma_\mathrm{0} \neq 1$, the 4-parameter generalized entropy deviates from the Bekenstein-Hawking entropy and consequently the present time Hubble parameter in our model seems to differ than the $\Lambda$CDM case. This may indicate a possible resolution of the Hubble tension in the present context of generalized entropic cosmology.
\end{itemize}

\section{Data analysis and results}\label{results}
Data analysis is a salient area of study in cosmology where we calculate the optimal value of the model parameters using different observational datasets. This helps us to obtain more accurate and persistent results. In this section we present a brief description of the various observational datasets used and the methodology adopted to constrain the parameters of the model. 
%For this analysis, we have used multiple sets of observational data, like the Hubble data \cite{solanki2021cosmic}, Cosmic Chronometer data (CC) \cite{singh2023new} and the most recent Pantheon dataset \cite{scolnic2018complete}. By fitting the parameters of the proposed model to these observational datasets, we determine their best-fit values which provides a more precise cosmological framework.\\
We have  performed the standard Bayesian analysis \cite{padilla2021cosmological} to obtain the posterior distribution of the parameters by employing a Markov Chain Monte Carlo (MCMC) method. For this we have used the publicly available emcee library package in Python \cite{foreman2019emcee} to carry out the MCMC analysis and the GetDist package \cite{lewis2019getdist} has been used for statistical analysis. The following publicly available observational datasets have been used to constrain the model parameters:  
\begin{itemize}
\item {\bf{Cosmic Chronometer data:}} In Cosmic Chronometry, the Hubble parameter $H(z)$ at different redshifts is usually determined through two approaches: (i) by extracting $H(z)$ from line-of sight of BAO data \cite{Chuang_2016, Bautista_2017} and (ii) by estimating $H(z)$ via the method of differential age (DA) of galaxies \cite{Simon:2005, Ratsimbazafy_2017} which relies on the relation
$$H(z) = -\frac{1}{1+z} \frac{dz}{dt}$$
where $\frac{dz}{dt}$ is approximated by determining the time interval $\Delta t$ corresponding to a given $\Delta z$ \cite{Ratsimbazafy_2017}.  This method is straightforward and enables astronomers to infer the expansion rate of the universe at different cosmic epochs, offering insights into fundamental properties of the universe. 
    %In this work we have used the 57 data points of Hubble parameter measurements in the redshift range $0.07 \le z \le 2.36$ of which 31 points have been measured via the method of differential age (DA) and the remaining 26 through BAO and other methods \cite{mhamdi2024}. The list of data points of $H(z)$ measurements in the redshift range $0.07 \le z \le 2.36$ can be found in \cite{Solanki_2021}.

\begin{table*}[t!]
\centering
\begin{tabular}{|c|c|c||c|c|c|}
\hline
\multicolumn{3}{|c||}{Non-correlated datapoints } & \multicolumn{3}{c|}{Correlated datapoints} \\
\hline
 $z$ & $H(z)$ & Reference & $z$ & $H(z)$ & Reference\\
 & $(km/s/Mpc)$ & & & $(km/s/Mpc)$ & \\
\hline
0.07 & 69 $\pm$ 19.6 & \cite{zhang2014four} & 0.1791 & 75 $\pm$ 4 & \cite{M_Moresco_2012}\\
0.09 & 69 $\pm$ 12 & \cite{Simon:2005} & 0.1993	& 75 $\pm$ 5 &\cite{M_Moresco_2012}\\
0.12 & 68.6 $\pm$ 26.2 & \cite{zhang2014four} & 0.3519	& 83 $\pm$ 14 & \cite{M_Moresco_2012}\\
0.17 & 83 $\pm$ 8 & \cite{Simon:2005} & 0.3802	& 83 $\pm$ 13.5 & \cite{Moresco_2016}\\
0.20 & 72.9 $\pm$ 29.6 & \cite{zhang2014four} & 0.4004 & 77 $\pm$ 10.2 & \cite{Moresco_2016}\\
0.27 & 77 $\pm$ 14 & \cite{Simon:2005} & 0.4247 & 87.1 $\pm$ 11.2 & \cite{Moresco_2016}\\
0.28 & 88.8 $\pm$ 36.6 & \cite{zhang2014four} & 0.4497 & 92.8 $\pm$ 12.9 & \cite{Moresco_2016} \\
0.40 & 95 $\pm$ 17 & \cite{Simon:2005} & 0.4783 & 80.9 $\pm$ 9 &  \cite{Moresco_2016}\\
0.47 & 89 $\pm$ 34 & \cite{Ratsimbazafy_2017} & 0.7812 & 105 $\pm$ 12 & \cite{M_Moresco_2012} \\
0.48 & 97 $\pm$ 60 & \cite{Simon:2005} & 0.5929 & 104 $\pm$ 13 & \cite{M_Moresco_2012} \\
0.75 & 98.8 $\pm$ 33.6 & \cite{Borghi_2022} & 0.6797 & 92 $\pm$ 8 & \cite{M_Moresco_2012}\\
0.8 & 113.1 $\pm$ 15.1 & \cite{Jiao_2023} & 0.8754 & 125 $\pm$ 17 & \cite{M_Moresco_2012}\\
0.88 & 90 $\pm$ 40 & \cite{Daniel_Stern_2010} & 1.037 & 154 $\pm$ 20 & \cite{M_Moresco_2012}\\
0.9 & 117 $\pm$ 23 & \cite{Daniel_Stern_2010} & 1.363 & 160 $\pm$ 33.6 & \cite{MorescoMNRAS2015}\\
1.26 & 135 $\pm$ 65 & \cite{Tomasetti_2023} & 1.965 & 186.5 $\pm$ 50.4 & \cite{MorescoMNRAS2015} \\ 
1.3 & 168 $\pm$ 17 & \cite{M_Moresco_2012} & & &\\
1.43 & 177 $\pm$ 18 & \cite{Simon:2005} & & &\\
1.53 & 140 $\pm$ 14 & \cite{Daniel_Stern_2010} &  & &\\
1.75 & 202 $\pm$ 40 & \cite{Daniel_Stern_2010} &  & &\\ 

\hline
\end{tabular}
\caption{$H(z)$ data corresponding to Cosmic Chronometer measurements along with corresponding uncertainties incorporating both systematic and statistical factors \cite{Moresco_2020}. The non-correlated data points are taken from various surveys while the correlated data points involves systematic uncertainties as discussed in \cite{Moresco_2020}.}
\label{CCdata}
\end{table*}
However, recent studies suggest that addressing systematic uncertainties is crucial for precise $H(z)$ measurements. Moresco et al. \cite{Moresco_2020, M_Moresco_2012, Moresco_2015, Moresco_2016} conducted a detailed analysis of the components contributing to these uncertainties.
Their research provides a detailed analysis of the factors contributing to these uncertainties, identifying four primary sources of systematic uncertainty: the stellar population synthesis (SPS) model, the metallicity of the stellar population, the star formation history (SFH) and residual star formation from a young, subdominant stellar component. The covariance matrix between the Hubble parameter values at redshifts $z_i$ and $z_j$, represented by $C_{ij}$, associated with the CC method can be expressed as \cite{Moresco_2020}
\begin{equation}\label{corr_cov}
C_{ij} = C_{ij}^{stat} + C^{young}_{ij} + C^{model}_{ij} + C^{met}_{ij}
\end{equation}
where $C_{ij}^{stat}$ is the contribution to the covariance matrix due to statistical errors, $C^{young}_{ij}$ is the contribution due to residual star formation from a young subdominant component, $C^{model}_{ij}$ is contribution due to the dependence on the chosen model and $C^{met}_{ij}$ is the contribution due to the  uncertainty in the estimate of the stellar metallicity. In this study, we have considered 34 data points from both correlated and non-correlated datasets in the redshift range $0.07$ to $1.965$, as listed in table \ref{CCdata}.
The $\chi^2$ function  for the cosmic chronometer dataset is defined as 
\begin{eqnarray}
\chi^2_{CC/non-corr} = \Delta \mathscr{A} C_1^{-1} \Delta \mathscr{A}^T \\
\chi^2_{CC/corr} = \Delta \mathscr{A} C_2^{-1} \Delta \mathscr{A}^T
\end{eqnarray}
 The vector $\mathscr{A}$ represents the 
 data points, such that each element is computed as
\begin{equation}
\Delta \mathscr{A}_i=[{H}^{th}(z_{i}) - {H}^{obs}(z_{i})]_{1 \times k}  
\end{equation}
$C_1^{-1}$ is the inverse of the covariance matrix for uncorrelated data points, given by, $C_1^{-1} = \left[\frac{1}{\sigma_{H}^2 (z_i)}\right]_{k \times k}$ where $\sigma_{H} (z_i)$ indicates the error associated with each data point in the CC data.  
The superscripts {\it obs} and {\it th} refer to the observational values and the corresponding theoretical values respectively and $\sigma_{H}$ indicates the error for each data point. Similarly, $C_2^{-1}$ is the inverse of the covariance matrix for correlated data points as given in equation (\ref{corr_cov}). The total $\chi^2$ for the CC dataset comprising of both correlated and uncorrelated datapoints is given by 
\begin{equation}
\chi^2_{CC} = \chi^2_{CC/non-corr} + \chi^2_{CC/corr}
\end{equation}

\item {\bf\textcolor{blue}{PantheonPlus + SH0ES data:}}
We have also used the comprehensive PantheonPlus compilation which comprises 1701 light curves from 1550 SNe, spanning a redshift range $0.001 \le z \le 2.2613$, sourced from 18 different studies \cite{Riess_2022,Brout_2022,Scolnic_2022}. Out of the 1701 light curves in the dataset, 77 of them are associated with galaxies that contain Cepheids. The PantheonPlus + SH0ES compilation introduces significant enhancements and primarily features an expanded sample size, particularly for SNe at redshifts below $0.01$. Additionally, notable improvements have been made in addressing systematic uncertainties associated with redshifts, intrinsic scatter models, photometric calibration, and peculiar velocities of Supernovae.\\
The definition of the $\chi^2$ function for PantheonPlus + SH0ES  dataset is given by \citep{scolnic2018complete}
\begin{equation}
\chi^{2}_{PantheonPlus+SH0ES}=\sum_{i,j=1}^{1701}(\mu^{th}-\mu^{obs})_{i} {C_{ij}}^{-1} (\mu^{th}-\mu^{obs})_{j}\\ 
\end{equation}
where  $\mu^{obs}$, $\mu^{th}$ represent the observed distance modulus and the corresponding theoretical value respectively and $C^{-1}$ corresponds to the inverse of covariance matrix for PantheonPlus + SH0ES sample. The theoretical distance modulus can be obtained as
\begin{equation}
\mu_{th}(z)=5log_{10}\frac{d_{L}(z)}{1 Mpc}+25,
\end{equation}
 where $d_{L}(z)$ denotes the luminosity distance which can be evaluated by integrating the expression
 \begin{equation}
d_{L}(z,\theta_{s}) = (1+z)\int_{0}^{z}\frac{dz'}{H(z',\theta_{s})},
 \end{equation}
where $H(z)$ is the expression of the Hubble parameter for the cosmological model and $\theta_{s}$ is the parameter space of the cosmological model.\\
\item{\bf\textcolor{blue}{DESI Data:}}
We have also used 12 DESI-BAO DR1 measurements \cite{desicollaboration2024desi2024vicosmological} across the redshift range $0.1 < z < 4.2$, including volume averaged distance $D_V (z)/r_d$, angular diameter distance $D_M(z)/r_d$, and comoving Hubble distance $D_H(z)/r_d$,
where $r_d$ is the sound horizon at the drag epoch. The effective redshift and the corresponding $D_M / r_d$, $D_H / r_d$, and $D_V / r_d$ ratios used in this work have been taken from table 1 of reference ~\cite{desicollaboration2024desi2024vicosmological}.
The $\chi^2$ function for BAO dataset is defined as 
\begin{equation}
\chi_{BAO}^2=X^TC^{-1}X
\end{equation}
where $X$ shows the transformation matrix and $C^{-1}$ is the inverse covariance matrix which are constructed using different functional form of $\frac{d_{A}(z_{*})}{D_{v}(z_{BAO})}$ (see \citep{desicollaboration2024desi2024vicosmological,roy2024dynamicaldarkenergylight} for details).\\

\item{\bf\textcolor{blue}{Compressed Planck likelihood data:}}
In this analysis, we utilize the Planck compressed likelihood data (referred to as P18) following the methodology outlined in \cite{Arendse:2019hev}. This compressed likelihood is advantageous as this provides the same constraining accuracy as full Planck likelihood. This approach employs the baryon physical density $\omega_b = \Omega_b h^2$ and the two shift parameters $\theta_*=r_s\left(z_{\text {dec }}\right) / D_A\left(z_{\text {dec }}\right), \quad \mathcal{R}=\sqrt{\Omega_M H_0^2} D_A\left(z_{\text {dec }}\right),
$ where, the redshift decoupling is $z_{d e c}$, and $D_A$ is the comoving angular diameter distance. The mean values of the above parameters are chosen to be $100 \omega_b=2.237\pm 0.015,100 \theta_s=1.0411\pm0.00031$, and $\mathcal{R}=1.74998 \pm 0.004$, with the corresponding correlation matrix provided in Appendix A of ~\cite{Arendse:2019hev}.\\
 The total $\chi^2$ for these combined datasets
is given by
$$\chi^2_{total} =\chi^2_{CC} + \chi^2_{PantheonPlus + SH0ES} + \chi^2_{DESI} + \chi^2_{P18} $$

\end{itemize}
\begin{figure}[ht]
\begin{center}
\includegraphics[width=0.7\columnwidth]{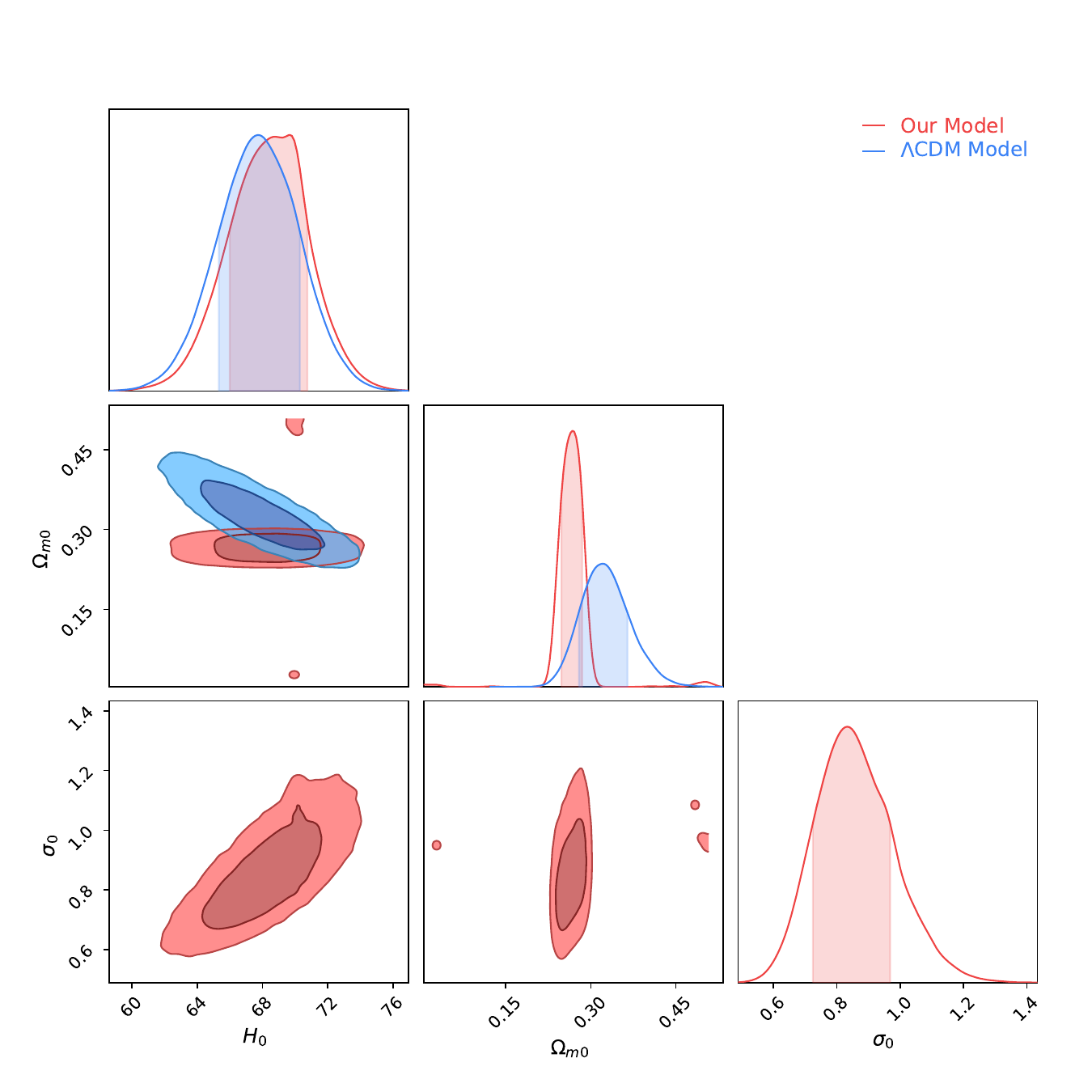}
\caption{The $1\sigma$ and $2\sigma$ likelihood contours for the model parameters of the generalized entropic DE model corresponding to CC dataset.} 
\label{cc3parameter}
\end{center}
\end{figure}
We perform the analysis considering $H_{0}$, $\Omega_{m0}$ and $\sigma_\mathrm{0}$ as free parameters keeping $\beta$ fixed at $\beta=1$. As mentioned earlier, for $\beta= \sigma_\mathrm{0} =1$, the proposed model behaves like a $\Lambda$CDM model. Our aim is to check the viability of the proposed generalized entropic DE model ensuring that its deviation from the $\Lambda$CDM model is minimum. There are two ways by which one can constrain the parameter space such that the deviation between the proposed model and the $\Lambda$CDM model is minimal: either one can set $\beta=1$ and vary $\sigma_\mathrm{0}$ or vice versa. It has been found that if we set $\beta=1$ and $\sigma_\mathrm{0} < 1$, the energy density for the entropic dark energy component remains positive throughout the evolution. However, if we set $\sigma_\mathrm{0}=1$ and vary $\beta$, the energy density corresponding to the entropic dark energy component becomes negative beyond $z \sim 3$. Recently dark energy models with negative cosmological constant \cite{Sen_2022, Adil_2023, Calder_n_2021, Menci_2024} or sign switching cosmological constant models \cite{Akarsu_2021, Sabogal:2025mkp, Akarsu:2024qsi} have been proposed where it has been demonstrated that the field evolves on a potential with an AdS minimum rather than the standard dS minimum. It has been shown that negative cosmological constant models have $\rho_{de} < 0$ in the early time resulting in a non-phantom equation of state for dark energy sector \cite{Sen_2022}; whereas at late times, $\rho_{de} > 0$, indicating a non-phantom to phantom transition of the universe. These models have been found to alleviate  cosmological tensions and provide a good fit to the observational data compared to the standard $\Lambda$CDM model. For the present work, we restrict ourselves to the case $\beta=1$ and $\sigma_\mathrm{0} < 1$, such that the energy density for the proposed entropic dark energy component remains positive. The evolution dynamics of the universe for the case where the energy density for the entropic dark energy component exhibits a transition from negative to positive value will be considered in a separate study.
%------------------------------------
\begin{figure}[ht]
\begin{center}
\includegraphics[width=0.49\columnwidth]{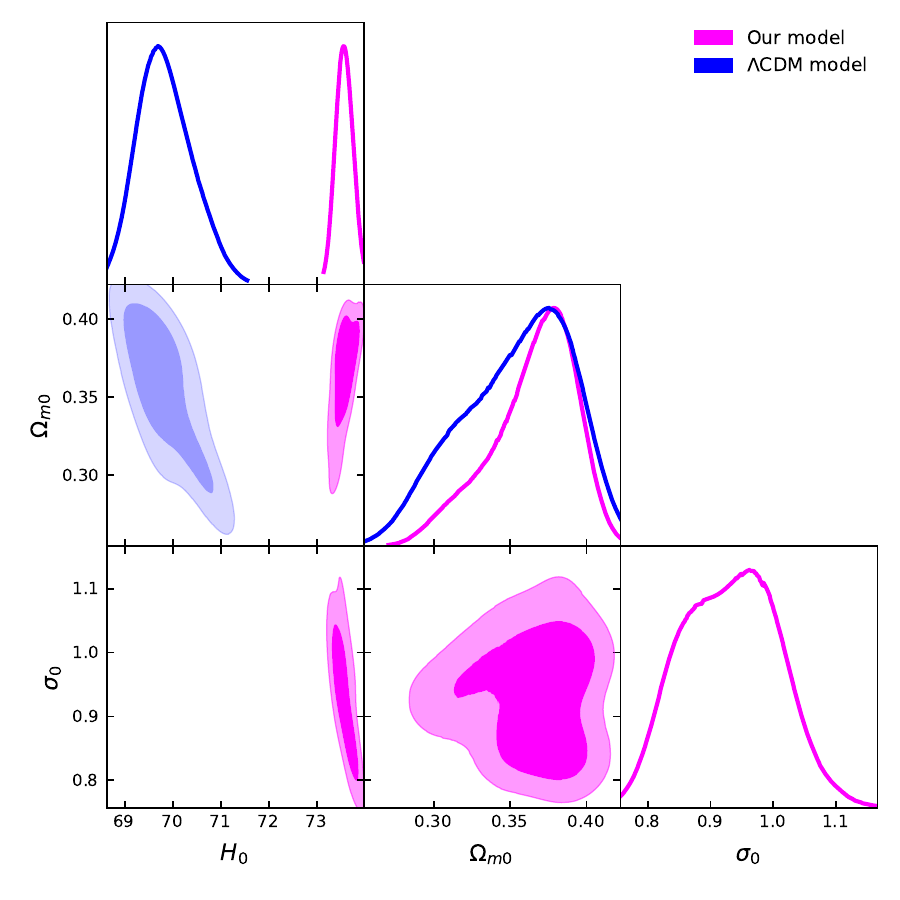}
\includegraphics[width=0.49\columnwidth]{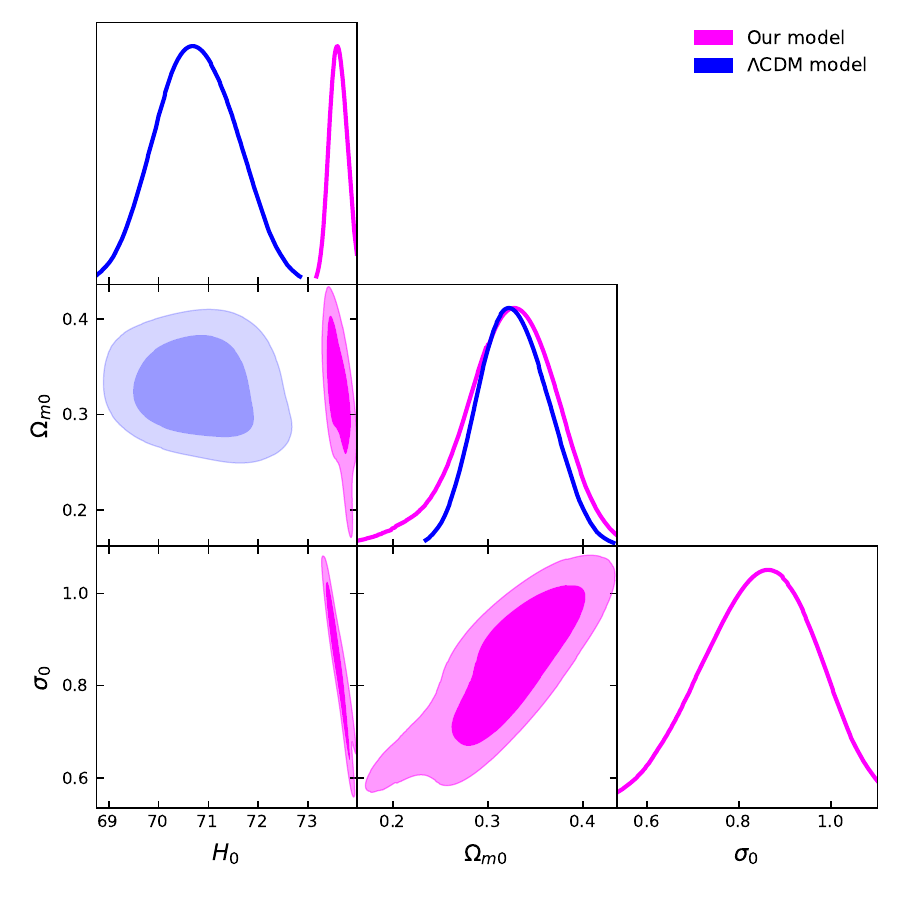}
\caption{The $1\sigma$ and $2\sigma$ iso-likelihood contours for generalized entropy constraining model parameters for PantheonPlus + SH0ES dataset (left panel) and CC+PantheonPlus+SH0ES datasets (right panel) with $\beta=1$.} 
\label{hcc+pan}
\end{center}
\end{figure}
%-------------------------------------
\begin{figure}[ht]
\begin{center}
\includegraphics[width=0.7\columnwidth]{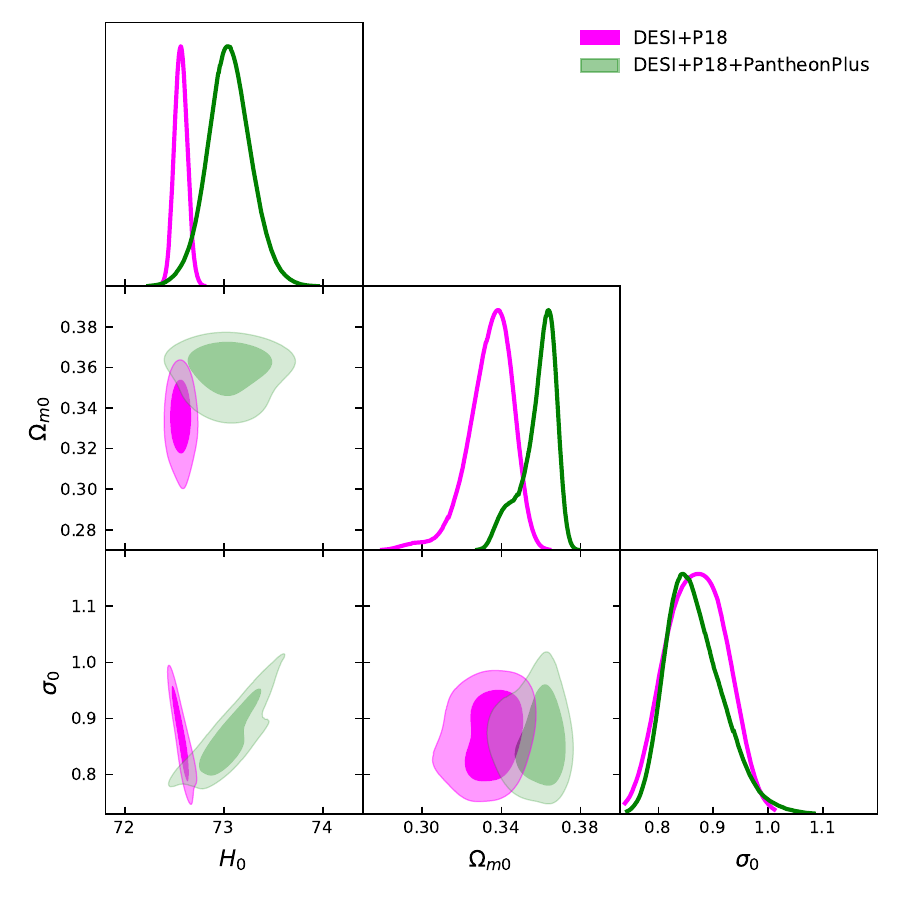}

\caption{The $1\sigma$ and $2\sigma$ contours for $H_0$, $\Omega_{m0}$ and $\sigma_0$ for DESI DR1 + P18 and DESI DR1 + P18 + PantheonPlus datasets with $\beta=1$.} 
\label{desi+p18+pan}
\end{center}
\end{figure}
%------------------------------------
For the MCMC analysis, we have chosen the flat priors on the base cosmological parameters as follows:
$H_0 \in [40, 120]$, $\Omega_{m0} \in [0.1, 1]$ and $\sigma_\mathrm{0} \in [0.2, 2.1]$. Figure \ref{cc3parameter} represents the $1\sigma$ and $2\sigma$ likelihood contours for the proposed generalized entropic dark energy model corresponding to CC dataset. As mentioned earlier, the CC dataset consists of data points from both correlated and non-correlated datasets in the redshift range $0.07 \le z \le  1.965$ where the correlated dataset take into account the covariance matrix between the data points. The best-fit values for the model parameters are obtained as $H_0 = 69.94^{+0.810}_{-3.95}$ km/s/Mpc, $\Omega_{m0}=0.272^{+0.013}_{-0.024}$ and $\sigma_\mathrm{0}=0.83^{+0.13}_{-0.11}$. We have also compared the model with the standard $\Lambda$CDM model for which the best-fit values are obtained as $H_0 = 67.9^{+2.4}_{-2.5}$ km/s/Mpc and $\Omega_{m0}=0.324^{+0.041}_{-0.045}$.  It is found that for the CC dataset, the value of $H_0$ is closer to the Planck results. We have also performed the analysis  considering the PantheonPlus + SH0ES, BAO datasets from the recent DESI DR1 collaboration along with the Compressed Planck likelihood data and the results have been presented in figure \ref{hcc+pan} and figure \ref{desi+p18+pan}. The resulting best-fit values of the model parameters have been tabulated in table \ref{table1}.
%%%%%%%%%%%%%%%%%%%%%%%%%%%%%
\begin{table}
\begin{center}
\begin{tabular}{ |c|c|c|c|c| } 
\hline
\hline
Dataset & Model & $H_0$ & $\Omega_{m0}$ & $\sigma_\mathrm{0}$ \\
& & (km/s/Mpc) & & \\
\hline
$\multirow{2}{*}{CC}$ & Our model & $69.94^{+0.81}_{-3.95}$  & $0.272^{+0.013}_{-0.024}$ & $0.83^{+0.13}_{-0.11}$  \\
\cline{2-5}
& $\Lambda$CDM & $67.9^{+2.4}_{-2.5}$  & $0.324^{+0.041}_{-0.045}$ & - \\
\hline

\multirow{2}{*}{PantheonPlus + SH0ES} & Our model & $73.34^{+0.119}_{-0.119}$ & $0.32^{+0.049}_{-0.033}$ & $0.89^{+0.035}_{-0.061}$ \\
\cline{2-5}
& $\Lambda$CDM & $69.06^{+0.477}_{-0.181}$  & $0.36^{+0.029}_{-0.037}$ & - \\
\hline
 
$\multirow{2}{*}{CC + PantheonPlus + SH0ES}$ & Our model & $73.24^{+0.155}_{-0.205}$  & $0.35^{+0.043}_{-0.046}$ & $0.93^{+0.09}_{-0.087}$ \\
\cline{2-5}
& $\Lambda$CDM & $68.49^{+0.016}_{-0.047}$  & $0.32^{+0.034}_{-0.029}$ & - \\
\hline
 
$\multirow{2}{*}{DESI + P18}$ & Our model & $72.56^{+0.049}_{-0.057}$  & $0.34^{+0.007}_{-0.012}$ & $0.87^{+0.055}_{-0.061}$ \\
\cline{2-5}
& $\Lambda$CDM & $68.48^{+1.163}_{-0.366}$  & $0.38^{+0.015}_{-0.034}$ & - \\
\hline
$\multirow{2}{*}{DESI + P18 + PantheonPlus + SH0ES}$ & Our model & $73.04^{+0.23}_{-0.184}$  & $0.36^{+0.004}_{-0.011}$ & $0.86^{+0.065}_{-0.042}$ \\
\cline{2-5}
& $\Lambda$CDM & $69.39^{+0.014}_{-0.007}$  & $0.38^{+0.017}_{-0.017}$ & - \\
\hline
\end{tabular}
\end{center}
\caption{Best-fit values of $H_0$, $\Omega_{m0}$ and $\sigma_\mathrm{0}$ for the generalized entropic DE model corresponding to various datasets.} 
\label{table1}
\end{table}
%---------------------------------------
As evident from table \ref{table1}, for the PantheonPlus + SH0ES dataset, the best-fit value of $H_0$ comes out to be $H_0=73.34^{+0.119}_{-0.119}$ km/s/Mpc which is compatible with the SH0ES result \cite{Sh0ES2019, H0LiCOW2019}. For the combined CC + PantheonPlus + SH0ES dataset or DESI + P18 + PantheonPlus + SH0ES datasets also, the best-fit value of $H_0$ comes out to be $\sim 73$ km/s/Mpc which is also higher than the Planck value and compatible with the SH0ES results. In order to check whether the high value of $H_0$ obtained for the proposed model is an artifact of SH0ES calibration or not, we have carried out the analysis for DESI + P18 and DESI + P18 + PantheonPlus + SH0ES datasets separately. As evident from figure \ref{desi+p18+pan} as well as table \ref{table1}, for both the datasets, the value of $H_0$ is closer to the SH0ES result. For DESI + P18 dataset, we obtain $H_0 =  72.56^{+0.049}_{-0.057}$ km/s/Mpc whereas when we include PantheonPlus dataset, the value of $H_0$ becomes $H_0 =73.04^{+0.23}_{-0.184}$ km/s/Mpc. Thus the results obtained indicate that the proposed generalized entropic dark energy model is capable of providing a possible resolution to the Hubble tension problem.
%---------------------------------------
\begin{widetext}
\begin{figure}[h!]
\caption{Comparison of the generalized entropic dark energy model with the $\Lambda$CDM model for CC and PantheonPlus + SH0ES dataset.}\label{h_vs_z}
%\begin{centering}
\subfigure[~Hubble parameter profile for the generalized entropic dark energy model. The black and red curves represent the Hubble function for the proposed model and the $\Lambda$CDM model respectively for the best-fit values corresponding to CC data. The blue dots with bars represent the CC dataset with error bar.]
{\includegraphics[width=0.6\columnwidth]{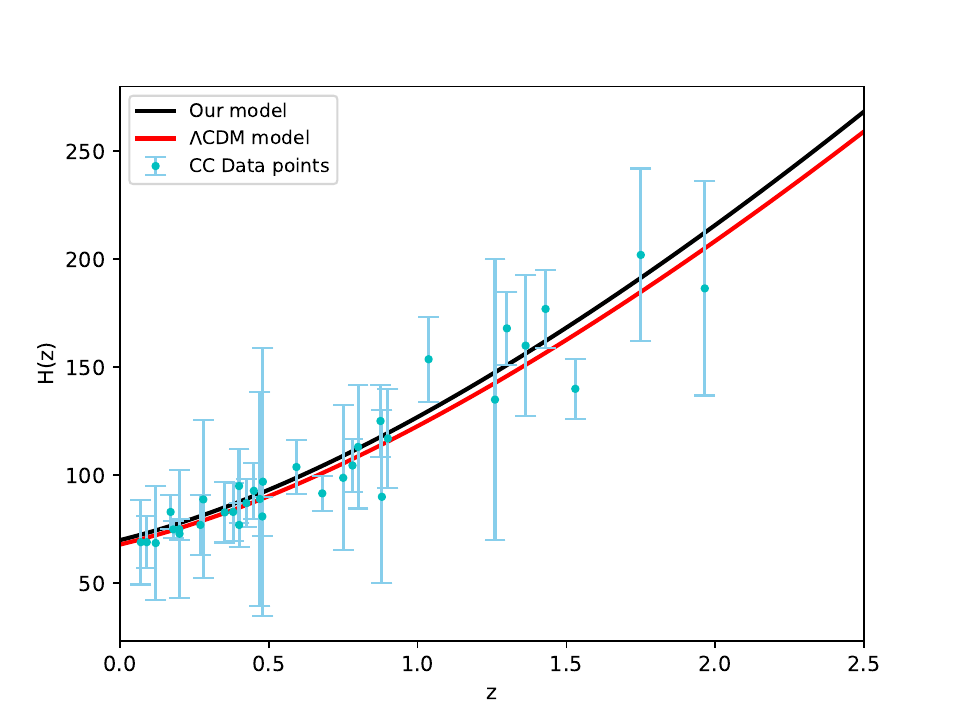}}
\subfigure[~Plot of distance modulus $\mu(z)$ for PantheonPlus + SH0ES dataset. The green dotted curve and the red curve represent the distance modulus function for the
proposed model and the $\Lambda$CDM model respectively.]
{\includegraphics[width=0.7\columnwidth]{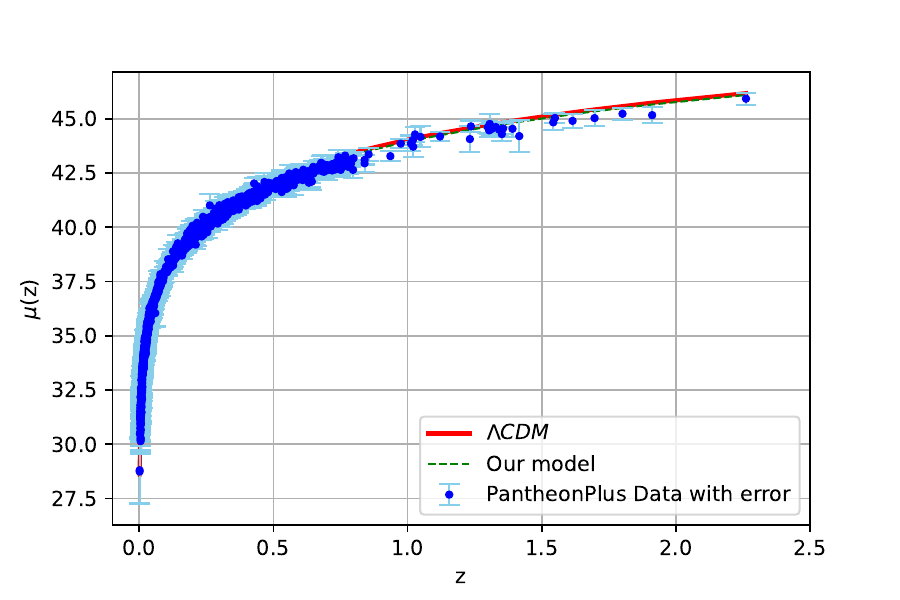}}
%\end{centering}
\end{figure}
\end{widetext}
%------------------------------------
%-------------------------------------
\begin{figure*}[ht]
\begin{center}
\includegraphics[width=0.49\columnwidth]{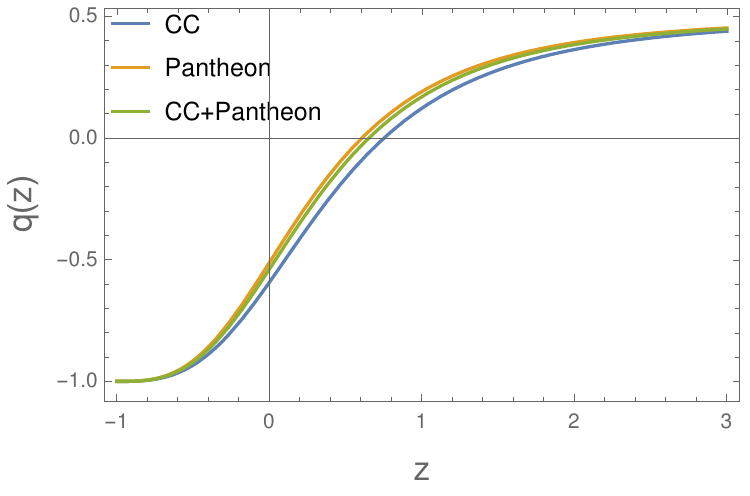}
\includegraphics[width=0.49\columnwidth]{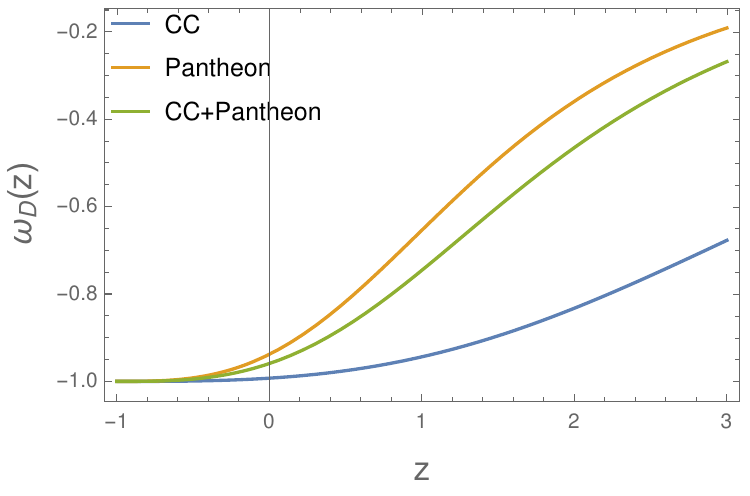}
\caption{Left Plot: Deceleration parameter, $q(z)$ vs. $z$ with the best-fitted parameters for different datasets. Right Plot: Dark energy EoS parameter, $\omega_\mathrm{D}(z)$ vs. $z$ with the best fitted parameters for different datasets.} 
\label{qvsz}
\end{center}
\end{figure*}
 %-------------------------------------
These interesting results motivated us to look into the evolution of few important cosmological parameters, like the Hubble parameter $H(z)$, the deceleration parameter $q(z)$, the equation of state parameter $\omega_D(z)$ etc. for the proposed model. These parameters provide a comprehensive set of quantities which can provide an insight to the important aspects of the cosmic evolution and can be used to test the consistency of the proposed model. Figure \ref{h_vs_z} represent the reconstructed Hubble parameter profile (figure 3(a)) and the distance modulus profile (figure 3(b)) for the proposed  model corresponding to CC and PantheonPlus + SH0ES datasets respectively. We have also plotted the profiles for the standard $\Lambda$CDM model for the sake of comparison and it is found that throughout the evolution, the Hubble parameter is in excellent agreement with the $\Lambda$CDM model and the observational datasets. 

The left plot of figure \ref{qvsz} represents the reconstructed deceleration parameter $q(z)$ given in equation~(\ref{deceleration parameter}) using the best-fit values listed in table \ref{table1}. It has been found that the deceleration parameter $q(z)$
undergoes a smooth transition from a decelerating to an accelerating phase, which depict the correct thermal history of the universe, consistent with the sequence of matter and dark energy dominated epochs. The transition redshift $z_t$ varies slightly for different datasets, but is close to the widely accepted value of $z \sim 0.5$ \cite{riess2001farthest, Muthukrishna_2016}. It has also been
observed that the model approaches $q = -1$ value in
far future ($z \rightarrow -1$) for all the datasets, which represents a future de-Sitter universe. 
%------------------------------------
%\begin{figure}[h!]
%\begin{center}
%\includegraphics[width=0.6\columnwidth]{eos-plot.pdf}
%\caption{Dark energy EoS parameter: $\omega_\mathrm{D}(z)$ vs. $z$ with the best fitted parameters for different datasets.} 
%\label{eosplot}
%\end{center}
%\end{figure}
%------------------------------------

We have also reconstructed the equation of state parameter for the DE component $\omega_D(z)$ using the best-fit values listed in table \ref{table1}. From the right plot of figure \ref{qvsz} it can be seen that for PantheonPlus + SH0ES dataset, the accelerating phase of the universe sets in around ($z_t \sim 2$) whereas the best-fit values corresponding to CC dataset indicates the onset of acceleration at much earlier epoch. All the datasets indicate that the proposed generalized entropic DE model will approach the $\Lambda$CDM value in far future ($z \rightarrow -1$).  
The evolution of these cosmological parameters indicate that the proposed model is capable of providing a viable evolutionary dynamics with a possible resolution to the Hubble tension problem.

\subsection{Statistical comparison of the model with the $\Lambda$CDM model}
As we have mentioned earlier, the values of the entropic parameters ensure that the deviation of the proposed model from the $\Lambda$CDM model is minimum. For a comprehensive analysis, we perform a statistical comparison of the proposed generalized entropic dark energy model with the standard $\Lambda$CDM model. For a given set of competing models, the compatibility and relative tension between the models can be determined utilizing the statistical tools Akaike Information Criterion (AIC) and Bayesian Information Criterion (BIC) \cite{Liddle_2007}.
The AIC and BIC estimators are calculated from the maximum log-likelihood $\left({\mathcal{L}}_{max} = exp\left(\frac{-\chi^2}{2}\right)\right)$ of the model and is given by \cite{ModAIC, Goswami:2024ymh}
\begin{eqnarray}
    AIC = -2~\mathrm{ln}~({\mathcal{L}}_{max}) + 2k  \\
BIC = -2~\mathrm{ln}~({\mathcal{L}}_{max}) + k~\mathrm{log}(N)   
\end{eqnarray}
where $N$ is the count of data points utilized for the sampling and $k$ is the number of independent model parameters. 
For comparison with the $\Lambda$CDM model, the relative difference
$\Delta AIC = (AIC_{model} - AIC_{\Lambda CDM})$ and $\Delta BIC = (BIC_{model} - BIC_{\Lambda CDM})$ are considered. According to Jeffrey’s scale \cite{Kass1995BayesF}, if $\Delta AIC \le 2$ this will indicate that the model is strongly favoured; if it falls in the range $4 < \Delta AIC < 7$, this indicates a mild tension between the models compared; if $\Delta AIC \ge 10$, it indicates no favoured evidence and corresponds to a strong tension \cite{Mandal_2023}. For BIC the
ranges are categorized as $\Delta BIC < 2$ corresponds to
strong evidence in favor of the model, $2 \le \Delta BIC \le 6$ indicates the moderate level, and $\Delta BIC > 6$ shows no evidence. In this work, we have compared our proposed model with the $\Lambda$CDM model and the relative $\Delta AIC$ and $\Delta BIC$ values have been listed in table \ref{tab:AICBIC}.
\begin{widetext}
\begin{table}[h!]
\centering

\begin{tabular}{|c|c|c|c|c|c|}
 \hline
  \hline
$\multirow{2}{*}{Dataset}$ & $\chi^2_{min}$  & AIC &$\Delta$AIC  & BIC & $\Delta$BIC\\ 
\cline{2-3}\cline{5-5}
 & Model \hspace{3mm} $\Lambda$CDM & Model \hspace{3mm} $\Lambda$CDM 
& & Model \hspace{3mm} $\Lambda$CDM& \\ 
 \hline
  CC & $-6.77$ \hspace{5mm} $-6.77$ & $-0.77$ \hspace{5mm} $-2.77$ 
   & $+2.0$ & $-2.18$ \hspace{5mm} $-3.71$ & $+1.53$ \\
  PantheonPlus + SH0ES & $-14.86$\hspace{5mm} $-14.77$ & $-8.86$ \hspace{3mm} $-10.77$  & $+1.91$  & $-5.17$\hspace{3mm} $-8.31$  & $+3.14$ \\
  CC + PantheonPlus & $-15.24$ \hspace{5mm} $-15.17$  & $-9.24$ \hspace{3mm} $-11.17$ &  $+1.93$ & $-5.52$ \hspace{3mm} $-8.69$ & $+3.17$\\
  + SH0ES & & & & & \\
 \hline
  \hline
\end{tabular}
\caption{The $AIC$, $BIC$ values for the proposed model in comparison with $\Lambda$CDM model and the corresponding relative differences $\Delta AIC$ and $\Delta BIC$}
\label{tab:AICBIC}
\end{table}
\end{widetext}
As noted from table \ref{tab:AICBIC}, for CC dataset, the proposed model and the 2 parameter $\Lambda$CDM model has the same value of $\chi^2_{min}$. But the $\Delta AIC$ and $\Delta BIC$ values indicate that the proposed generalised entropic dark energy model is strongly favored by the observational data. Further, from these $\Delta$BIC values one can infer that  there is a weak evidence in favor of the $\Lambda CDM$ model and the proposed model exhibits mild tension. 
\section{Concluding Remarks}\label{Sec-conclusion}
In this work, we have proposed a new candidate for dark energy arising from 4-parameter generalized entropy function $S_g$. This generalized entropy function generalizes all the known entropy functions proposed so far, for suitable representations of the entropic parameters. The generalized entropic dark energy is characterized by entropic energy density component ($\rho_g$) and entropic pressure component ($p_g$) which are defined in terms of the entropic parameters. Considering modified Friedmann equations corresponding to the generalized entropy function, the cosmological implications of the model has been studied. We have also tested the consistency of the model using latest observation datasets like Cosmic Chronometer, PantheonPlus + SH0ES, DESI-DR1 and compressed Planck likelihood datasets. For this we have carried out the $\chi^2$- minimization method and have performed the Markov Chain Monte Carlo (MCMC) analysis \citep{padilla2021cosmological} using emcee package \citep{foreman2019emcee}. It has been found that the proposed generalized entropic dark energy model can efficiently describe the late time cosmic acceleration preceded by a decelerated expansion phase. The deceleration parameter $q(z)$ shows a smooth transition from positive to negative value in recent past which is essential for the unhindered structure formation of the universe. Although the equation of state parameter $\omega_D$ exhibits deviation from the $\Lambda$CDM model at present epoch, it approaches $\omega_D = -1$
value in far future ($z \rightarrow -1$). {Most importantly, for PantheonPlus + SH0ES dataset or for DESI + P18 datasets, the MCMC analysis yields the present value of the Hubble parameter to be $H_0 \approx 73$ km/s/Mpc which is closer to the SH0ES result \cite{Sh0ES2019, H0LiCOW2019}. The AIC, BIC analysis and Jeffrey’s scale indicates that for CC data, the proposed model is strongly favored. For other datasets, although the $\Lambda CDM$ model is slightly favored by the observational datasets, the proposed generalized entropic model exhibits a mild tension but could address the Hubble tension problem.\\
Thus as a whole, the 4-parameter generalized entropic model serves as a viable dark energy candidate and concomitantly resolves the Hubble tension issue.\\

\vspace{8mm}

{\bf{Acknowledgements: }}
This work is partially supported by SERB, Government of India
through the project CRG/2023/000185. SD would like to acknowledge IUCAA, Pune for providing support through the associateship programme. The authors would like to acknowledge the support and facilities under ICARD, Pune at Department of Physics, Visva-Bharati, Santiniketan.

\bibliographystyle{apsrev4-1}
\bibliography{bibliography}
\end{document}